\documentstyle[preprint,aps,amstex,amssymb]{revtex}

\begin{document}
\draft

\title{Quantization of massive Abelian antisymmetric tensor field \\ 
and linear potential}

\author{Shinichi Deguchi\footnote{Email address:
deguchi@@phys.cst.nihon-u.ac.jp }and 
Yousuke Kokubo\footnote{Email address:
kokubo@@phys.cst.nihon-u.ac.jp }}
\address{Institute of Quantum Science, Nihon University \\
Tokyo 101-8308, Japan}

\maketitle

\begin{abstract}
We discuss a quantum-theoretical aspect of the massive Abelian antisymmetric 
tensor gauge theory with antisymmetric tensor current. 
To this end, an Abelian rank-2 antisymmetric tensor field is 
quantized both in the covariant gauge with an arbitrary gauge parameter 
and in the axial gauge of the Landau type. 
The covariant quantization yields the generating functional written in terms 
of an antisymmetric tensor current and its divergence. 
Origins of the terms in the generating functional are clearly understood 
in comparison with the quantization in the unitary gauge. 
The quantization in the axial gauge with a suitable axis 
directly yields the generating functional which is same as that obtained 
by using Zwanziger's formulation for electric and magnetic charges. 
It is shown that the generating functionals lead to 
a composite of the Yukawa and the linear potentials. 
\end{abstract}

\newpage

\section{Introduction}

Field theories that yield the linear potential 
are very important and attractive 
to particle physics, since those theories may be 
utilized to describe the confinement of quarks and gluons 
and furthermore might be considered to be effective theories of 
quantum chromodynamics (QCD). 
The dual Abelian Higgs model (DAHM), or the dual Ginzburg--Landau theory,  
\cite{Suz,SST} has well been known as 
one of such theories.

Recently, an extended dual Abelian Higgs model (EDAHM) has been proposed  
by Baker et al. \cite{BBDV} and by Antonov and Ebert \cite{AE} to 
incorporate external (color-)electric charges such as quarks into the DAHM. 
Using the path integral method \cite{Suga,ACPZ}, 
Antonov and Ebert have also derived a dual form of the Lagrangian that 
defines the EDAHM in the London limit.  
In the dual Lagrangian, namely the dual form of Lagrangian, 
the Abrikosov-Nielsen-Olesen (open) string in the EDAHM is represented by 
a vorticity tensor current with the non-zero divergence. 
The dual Lagrangian, which is written in terms of 
an Abelian rank-2 antisymmetric tensor field, an Abelian vector field, 
a vorticity tensor current and its divergence, 
has the exactly same form as the Lagrangian that Kalb and Ramond have 
discovered in the action-at-a-distance theory between open strings \cite{KR}. 
Kalb and Ramond showed that this Lagrangian describes a massive 
Abelian rank-2 antisymmetric tensor field coupled to an open string.

The present letter is intended to deal with a general rank-2 antisymmetric 
tensor current rather than the vorticity tensor current.  
Accordingly, the gauge theory found by Kalb and Ramond will be called 
{\it massive Abelian antisymmetric tensor gauge theory} (MAATGT) 
{\it with antisymmetric tensor current} (ATC) on condition that  
the vorticity tensor current is replaced by a general ATC.

The purpose of the present letter is to discuss a quantum-theoretical 
aspect of the MAATGT with ATC by performing quantization of 
an Abelian rank-2 antisymmetric tensor field contained in this theory. 
In Ref.\cite{AE}, Antonov and Ebert have actually considered the quantization 
of such a field. 
However, their procedure was complicated, because they carried out the 
quantization in the unitary gauge.  In contrast, we carry out 
the quantization in the covariant gauge with an arbitrary gauge parameter. 
By virtue of taking the covariant gauge, the generating functional written 
in terms of the ATC and its divergence is derived in a straightforward way.
Furthermore, origins of the terms constituting the generating functional 
are clearly understood. 
The quantization in the covariant gauge is also important to investigate 
properties of an effective gauge theory of SU(2) QCD that the present 
authors have recently studied \cite{DK}. 
The reason for this is that the effective gauge theory involves 
a covariantly gauge fixed version of the Lagrangian that defines 
the MAATGT with ATC.

In addition to the covariant quantization, we consider noncovariant 
quantization in the axial gauge of the Landau type \cite{BEMSSZ}. 
Choosing a suitable axis in the axial gauge condition, 
we can directly derive a generating functional written in terms of 
an Abelian vector field and the divergence of the ATC. 
This generating functional agrees with that found 
in the DAHM \cite{Suz,SST} constructed by using Zwanziger's 
formulation for electric and magnetic charges \cite{Zwan}.
The generating functionals obtained in the covariant and the noncovariant 
quantizations turn out to be the same for a specific form of the ATC. 
We shall see that these generating functionals lead to a composite of 
the Yukawa and the linear potentials.

The present letter is organized as follows. 
In Sec. 2, we introduce the MAATGT with ATC at the classical level. 
The covariant quantization is discussed in Sec. 3, and 
the noncovariant quantization in axial gauge is discussed in Sec. 4. 
Section 5 is devoted to a summary and discussion.

\section{Gauge theory of \\ massive Abelian antisymmetric tensor field}

Let $B_{\mu\nu}(=-B_{\nu\mu})$ be an Abelian antisymmetric tensor field and 
$A_{\mu}$ be an Abelian vector field. We consider the antisymmetric tensor 
gauge theory in four dimensions characterized by the Lagrangian 
\begin{eqnarray}
{\cal L}_{0} &=& {1\over12} H_{\mu\nu\rho} H^{\mu\nu\rho} 
-{1\over4}(mB_{\mu\nu}-F_{\mu\nu}) (mB^{\mu\nu}-F^{\mu\nu}) 
\nonumber \\ 
& &+{1\over2}mB_{\mu\nu}J^{\mu\nu}+A_{\mu}@,@,j^{\mu} \,,
\label{1}
\end{eqnarray}
%(1)
%
with 
\begin{align}
H_{\mu\nu\rho} &\equiv \partial_{\mu}B_{\nu\rho} +\partial_{\nu}B_{\rho\mu} 
+\partial_{\rho}B_{\mu\nu} \,, 
\label{2}
\\
F_{\mu\nu} &\equiv \partial_{\mu}A_{\nu}-\partial_{\nu}A_{\mu} \,,
\label{3}
\end{align}
%(2)(3)
%
where $m$ is a constant with dimension of mass, 
$J^{\mu\nu}(=-J^{\nu\mu})$ is an antisymmetric tensor current, and 
$j^{\mu}$ is a vector current satisfying 
\begin{align}
\partial_{\nu}J^{\nu\mu}&=j^{\mu}\,, 
\label{4}
\\
\partial_{\mu}@,@,@,j^{\mu}&=0 \,.
\label{5}
\end{align}
%(4)(5)
%
Equations (\ref{4}) and (\ref{5}) together guarantee that ${\cal L}_{0}$ 
remains invariant under the gauge transformation 
\begin{align}
\delta B_{\mu\nu}&=\partial_{\mu}\Lambda_{\nu}-\partial_{\nu}\Lambda_{\mu}\,,
\label{6}
\\
\delta A_{\mu}&=m\Lambda_{\mu}+\partial_{\mu}\lambda \,,
\label{7}
\end{align}
%(6)(7)
%
with gauge parameters $\Lambda_{\mu}$ and $\lambda$. 
The Lagrangian ${\cal L}_{0}$ describes an Abelian rank-2 antisymmetric tensor 
field with mass $m$, and we call the gauge theory defined by ${\cal L}_{0}$ 
{\it massive Abelian antisymmetric tensor gauge theory {\rm (MAATGT)} with 
antisymmetric tensor current {\rm (ATC)}.} 
In four dimensions, a massive rank-2 antisymmetric tensor field 
can be interpreted as a massive pseudovector field \cite{KR}.  
The sole physical component of $B_{\mu\nu}$ then corresponds to 
the longitudinal mode of a massive pseudovector field, 
while the two physical components of $A_{\mu}$ correspond to 
its transverse modes.

By using Eq. (\ref{4}), the Lagrangian ${\cal L}_{0}$ can simply be expressed 
as 
\begin{eqnarray}
{\cal L}_{0} ={1\over12} 
{\widetilde H}_{\mu\nu\rho} {\widetilde H}^{\mu\nu\rho} 
-{1\over4}m^{2}{\widetilde B}_{\mu\nu} {\widetilde B}^{\mu\nu}
+{1\over2}m{\widetilde B}_{\mu\nu}J^{\mu\nu}\,,
\label{8}
\end{eqnarray}
%(8)
%
where $\widetilde{B}_{\mu\nu}\equiv B_{\mu\nu}-m^{-1}F_{\mu\nu}$ and 
$\widetilde{H}_{\mu\nu\rho}\equiv \partial_{\mu}\widetilde{B}_{\nu\rho}
+\partial_{\nu}\widetilde{B}_{\rho\mu}
+\partial_{\rho}\widetilde{B}_{\mu\nu}@,@,$. 
Since $\widetilde{B}_{\mu\nu}$ is gauge invariant, the gauge invariance of 
${\cal L}_{0}$ is readily seen from Eq. (\ref{8}). 
It is also obvious from Eq. (\ref{8}) that ${\cal L}_{0}$ describes  
an Abelian rank-2 antisymmetric tensor field with mass $m$. 
In a manner of speaking, we can say that 
the massless field $B_{\mu\nu}$ develops into the massive field 
$\widetilde{B}_{\mu\nu}$ by eating the two physical degrees of freedom of 
$A_{\mu}@,@,$.  By analogy with the usual massive vector field theory, we may 
call Eq. (\ref{8}) the Lagrangian ${\cal L}_{0}$ in the unitary gauge. 
Antonov and Ebert have discussed the quantization of $\widetilde{B}_{\mu\nu}$ 
by carrying out the path-integration over $\widetilde{B}_{\mu\nu}$ in the 
generating functional defined with the Lagrangian in Eq. (\ref{8}) \cite{AE}. 

\section{Covariant Quantization}

In the present letter, starting from the generating functional defined with 
the Lagrangian in Eq. (\ref{1}), we first perform covariant quantization of 
$B_{\mu\nu}@,@,$, and we shall consider noncovariant quantization of 
${B}_{\mu\nu}$ in the axial gauge later. 
Both the covariant and the noncovariant quantizations will be done 
by using the path-integral method based on 
the Becchi-Rouet-Stora-Tyutin (BRST) formalism \cite{kimu,BEMSSZ}.

Noting the gauge transformation rule in Eq. (\ref{6}) and its reducibility, 
we now introduce the following ghost and auxiliary 
fields associated with $B_{\mu\nu\,}$: anticommuting vector fields 
$\rho_{\mu}$ and $\bar{\rho}_{\mu}@,@,$,  
a commuting vector field $\beta_{\mu}@,@,$, 
anticommuting scalar fields $\chi$ and $\bar{\chi}@,@,$, and commuting scalar 
fields $\sigma$, $\varphi$ and $\bar{\sigma}$. 
In addition, considering the gauge transformation rule in Eq. (\ref{7}), 
we introduce the following ghost and auxiliary fields associated 
with $A_{\mu\,}$: 
anticommuting scalar fields $c$ and $\bar{c}@,@,$, and commuting scalar field 
$b@,@,$. 
The BRST transformation $\boldsymbol{\delta}$ is defined for 
$B_{\mu\nu}$ and $A_{\mu}$ by replacing the gauge parameters $\Lambda_{\mu}$ 
and $\lambda$ in Eqs. (\ref{6}) and (\ref{7}) by the ghost fields  
$\rho_{\mu\,}$ and $c\,$: 
\begin{eqnarray}
\boldsymbol{\delta} B_{\mu\nu} &=&
\partial_{\mu}\rho_{\nu}-\partial_{\nu}\rho_{\mu} \,, 
\label{9} 
\\ 
\boldsymbol{\delta} A_{\mu}&=& m\rho_{\mu}+\partial_{\mu}c \,. 
\label{10}
\end{eqnarray}
%(9)(10)
%
The BRST transformation rules of the ghost and auxiliary fields are defined 
so as to satisfy the nilpotency condition $\boldsymbol{\delta}^{2}=0\,$:       
\begin{alignat}{2}
\boldsymbol{\delta}\rho_{\mu} &=-i\partial_{\mu}\sigma \,, &\qquad 
\boldsymbol{\delta}\sigma &=0 \,, 
\nonumber \\ 
\boldsymbol{\delta}\bar{\rho}_{\mu} &=i\beta_{\mu} \,, &\qquad 
\boldsymbol{\delta}\beta_{\mu} &=0 \,, 
\nonumber \\ 
\boldsymbol{\delta}\bar{\sigma} &=\bar{\chi} \,, &\qquad 
\boldsymbol{\delta}\bar{\chi}&=0 \,, 
\nonumber \\ 
\boldsymbol{\delta}\varphi &=\chi \,, &\qquad 
\boldsymbol{\delta}\chi &=0 \,, 
\label{11}
\\
& &\qquad & 
\nonumber \\
\boldsymbol{\delta}c &=im\sigma \,, 
\nonumber \\ 
\boldsymbol{\delta} \bar{c} &=ib \,, &\qquad 
\boldsymbol{\delta} b &=0 \,. 
\label{12}
\end{alignat}
%(11)(12)
%
To quantize $B_{\mu\nu}$ and $A_{\mu}$ in a covariant manner, we now take 
the following gauge-fixing terms \cite{kimu,DMN}:  
\begin{align}
{\cal L}_{\rm G1} &=
-i\boldsymbol{\delta} \bigg[ B_{\mu\nu}\partial^{\mu}\bar{\rho}^{\nu}
-{k\over2}\beta_{\mu}\bar{\rho}^{\mu} \bigg] @,@,, 
\label{13}
\\
{\cal L}_{\rm G2} &=
i\boldsymbol{\delta} \big[ \,\rho^{\mu}\partial_{\mu}\bar{\sigma}
+\bar{\rho}^{\mu}(\partial_{\mu}\varphi+uA_{\mu}) \big] \,, 
\label{14}
\\
{\cal L}_{\rm G3} &=
i\boldsymbol{\delta} \Big[ A^{\mu} \partial_{\mu}\bar{c} 
-\frac{\alpha}{2} \,\bar{c}\,b \,\Big] @,@,,
\label{15}
\end{align}
%(13)(14)(15) 
%
where $k$ and $\alpha$ are gauge parameters, and $u$ is a parameter with 
dimension of mass. 
Owing to the nilpotency of $\boldsymbol{\delta}$, each of these gauge-fixing 
terms is invariant under the BRST transformation. 
The first term ${\cal L}_{\rm G1}$ explicitly breaks  
the invariance of ${\cal L}_{0}$ under the gauge transformation generated  
by $\Lambda_{\mu}@,@,$, i.e. $\delta B_{\mu\nu}=\partial_{\mu}\Lambda_{\nu}
-\partial_{\nu}\Lambda_{\mu}@,@, , \; \delta A_{\mu}=m\Lambda_{\mu}@,@,$.  
The second term ${\cal L}_{\rm G2}$ is required to break 
the invariance of ${\cal L}_{\rm G1}$ under the {\it secondary} gauge 
transformation $\delta\rho_{\mu}=\partial_{\mu}\varepsilon@,@, , \;
\delta\bar{\rho}_{\mu}=\partial_{\mu}\bar{\varepsilon}@,@,$, 
with anticommuting gauge parameters $\varepsilon$ and $\bar{\varepsilon}$. 
The third term ${\cal L}_{\rm G3}$ is a gauge-fixing term that breaks  
the invariance of ${\cal L}_{0}$ and ${\cal L}_{\rm G2}$ under the 
gauge transformation $\delta A_{\mu}=\partial_{\mu}\lambda@,@, , \;
\delta \varphi=-u\lambda@,@,$. 
The two terms ${\cal L}_{\rm G1}$ and ${\cal L}_{\rm G2}$ are necessary to 
quantize $B_{\mu\nu}@,$ and the associated ghost fields 
$\rho_{\mu}$ and $\bar{\rho}_{\mu}@,$, 
while ${\cal L}_{\rm G3}$ is necessary to quantize $A_{\mu}@,$. 
Carrying out the BRST transformation in the right hand sides of 
Eqs. (\ref{13}), (\ref{14}) and (\ref{15}), we obtain 
\begin{eqnarray}
& & {\cal L}_{\rm G1}+{\cal L}_{\rm G2}+{\cal L}_{\rm G3} 
\nonumber \\
&=& -\beta^{\nu}(\partial^{\mu}B_{\mu\nu}+\partial_{\nu}\varphi
+uA_{\nu})-{k\over2}\beta_{\mu}\beta^{\mu}
\nonumber \\
& &
-i\bar{\rho}^{\nu} \{(\square+um)\rho_{\nu}
-\partial_{\nu}\partial^{\mu}\rho_{\mu}
+\partial_{\nu}\chi 
+u\partial_{\nu}c \}
\nonumber \\
& &
-i\rho^{\nu}(\partial_{\nu}\bar{\chi}-m\partial_{\nu} \bar{c})
-\bar{\sigma}@, \square @,
\sigma +i\bar{c}@,@,@,@, \square @,c 
+{\cal L}_{b} 
\nonumber \\
& & 
+\,\mbox{total derivative}\,,
\label{16}
\end{eqnarray} 
%(16)
%
where $\square \equiv\partial_{\mu}\partial^{\mu}$ and 
\begin{eqnarray}
{\cal L}_{b}\equiv b@,@,@,@,@, \partial^{\mu}A_{\mu} 
+\frac{\alpha}{2} @,@,@,@,@, b^{2} \,. 
\label{17}
\end{eqnarray}
%(17)
%

Let us now consider the generating functional 
\begin{eqnarray}
Z[J^{\mu\nu}, j^{\mu}@,]= N_{0} \int {\frak D}@!@!{\cal M} 
\exp\!\left[\,i\int d^{4}x ({\cal L}_{0}+
{\cal L}_{\rm G1}+{\cal L}_{\rm G2}+{\cal L}_{\rm G3}) \right] ,
\label{18}
\end{eqnarray}
%(18)
%
with the path-integral measure 
\begin{eqnarray}
{\frak D}@!@!{\cal M}
\equiv {\frak D}@!@!B_{\mu\nu}@,@,@,@,@, {\frak D}@!@!A_{\mu}@,@,@,@,@, 
{\frak D}\rho_{\mu}@,@,@,@,@, {\frak D}\bar{\rho}_{\mu}@,@,@,@,@,
{\frak D}\beta_{\mu}@,@,@,@,@,
{\frak D}\chi@,@,@,@,@,@,@,{\frak D}\bar{\chi}@,@,@,@,@,@,@,@, 
{\frak D}\sigma@,@,@,@,@,@,@, {\frak D}\bar{\sigma}@,@,@,@,@,@,@, 
{\frak D}\varphi@,@,@,@,@,@,@,
{\frak D}c\, {\frak D}\bar{c}\, {\frak D}b \,.
\label{19}
\end{eqnarray}
%(19)
%
Here and hereafter, $N_{i}$ $(@,@,@,i=0,1,2,3,4,5,6 @,@,@,)$ denote 
constants. We first note that the integrations over $\chi$ and 
$\bar{\chi}$ in Eq. (\ref{18}) yield 
the delta functions $\prod_{x}\delta(\partial^{\nu}\bar{\rho}_{\nu})$ and 
$\prod_{x}\delta(\partial^{\nu}\rho_{\nu})$. 
These functions enable us to remove the three terms 
$i\bar{\rho}^{\nu}\partial_{\nu}\partial^{\mu}\rho_{\mu}@,@,$, 
$-iu\bar{\rho}^{\nu}\partial_{\nu}c$  
and $im\rho^{\nu}\partial_{\nu}\bar{c}$ from the exponent of Eq. (\ref{18}). 
After removing them, we express the delta functions  
$\prod_{x}\delta(\partial^{\nu}\bar{\rho}_{\nu})$ and  
$\prod_{x}\delta(\partial^{\nu}\rho_{\nu})$ in the form of the integrals 
over $\chi$ and $\bar{\chi}$ again. 
Then, the integration over $\rho_{\mu}$ and $\bar{\rho}_{\mu}$  
yields 
$\{ \det(\square+um) \}^{4} \exp \! \left[\, i\int d^{4}x \, 
\bar{\chi} \{ i @,@,@, \square@,@,(\square+um)^{-1} \} \chi \right]$, and 
the integration over $\chi$ and $\bar{\chi}$ to be done next gives 
$\det\{@, \square@,@,(\square+um)^{-1} \}$. The integration over 
$\sigma$ and $\bar{\sigma}$ can be carried out immediately to get  
$(\det\square)^{-1}@,$, while that over $c$ and $\bar{c}$ can be carried out 
to get $\det\square@,$. 
As a consequence, the generating functional $Z$ is written as 
\begin{eqnarray}
Z[J^{\mu\nu}, j^{\mu}@,]&=& 
N_{1} \int{\frak D}@!@!B_{\mu\nu}@,@,@,@,@,  {\frak D}@!@!A_{\mu}@,@,@,@,@, 
{\frak D}\beta_{\mu}@,@,@,@,@, {\frak D}\varphi@,@,@,@,@,@,@, {\frak D}b 
\nonumber \\
& & \times \exp \biggl[ \, i \int d^{4}x \biggl\{ 
-{1\over4}B_{\mu\nu}(\square+m^{2}) B^{\mu\nu}
\biggr. \biggr.
\nonumber \\
& & 
-{1\over2}\partial^{\mu}B_{\mu\nu}\partial_{\rho}B^{\rho\nu} 
+\frac{m}{2} B_{\mu\nu}F^{\mu\nu} -{1\over4}F_{\mu\nu}F^{\mu\nu} 
\nonumber \\
& & -\beta^{\nu}(\partial^{\mu}B_{\mu\nu}+\partial_{\nu}\varphi
+uA_{\nu})-{k\over2}\beta_{\nu}\beta^{\nu} +{\cal L}_{b}
\nonumber \\
& &
\biggl. \Bigl.
+{1\over2}mB_{\mu\nu}J^{\mu\nu}+A_{\mu}@,@,j^{\mu} 
\biggr\} \biggr] \;.
\label{20}
\end{eqnarray}
%(20)
%

We now suppose $k\neq0@,@,$. In this case, 
the integration over $\beta_{\mu}$ in 
Eq. (\ref{20}) reduces to a Gaussian integration and leads to 
\begin{eqnarray}
Z[J^{\mu\nu}, j^{\mu}@,]&=& 
N_{2} \int{\frak D}@!@!B_{\mu\nu}@,@,@,@,@, {\frak D}@!@!A_{\mu}@,@,@,@,@,
{\frak D}\varphi@,@,@,@,@,@,@,
{\frak D}b 
\nonumber \\
& & \times \exp \biggl[ \, i \int d^{4}x \biggl\{ 
-{1\over4}B_{\mu\nu}(\square+m^{2}) B^{\mu\nu}
\nonumber \\
& &
+{1\over2}\bigg( {1\over k}-1 \bigg)
\partial^{\mu}B_{\mu\nu}\partial_{\rho}B^{\rho\nu} 
\biggr. \biggr.
\nonumber \\
& & 
+{1\over2}B_{\mu\nu}\biggl( \Bigl(m-\frac{u}{k}\Bigr) F^{\mu\nu}
+mJ^{\mu\nu} \bigg) \biggr.
-{1\over4}F_{\mu\nu}F^{\mu\nu} 
\nonumber \\
& & \biggl. \biggl.
+{1\over2k}(\partial_{\mu}\varphi+uA_{\mu})
 (\partial^{\mu}\varphi+uA^{\mu})+{\cal L}_{b}+A_{\mu}@,@,j^{\mu} 
\biggr\} \biggr]\,. 
\label{21}
\end{eqnarray}
%(21)
%
The integration over $B_{\mu\nu}$ in Eq. (\ref{21}) is carried out in the 
same way as one uses to derive the free propagator of $B_{\mu\nu}@,@,$.
The result reads
\renewcommand{\thefootnote}{\fnsymbol{footnote}}
\begin{eqnarray}
Z[J^{\mu\nu}, j^{\mu}@,]&=& 
N_{3} \int{\frak D}@!@!A_{\mu}@,@,@,@,@, {\frak D}\varphi@,@,@,@,@,@,@, 
{\frak D}b 
@,@, \exp \biggl[ \, i \int d^{4}x \biggl\{ 
-{1\over4}F_{\mu\nu}F^{\mu\nu} 
\biggr. \biggr.
\nonumber \\
& & 
+{1\over8} \biggl( \Bigl(m-\frac{u}{k}\Bigr) F_{\mu\nu}+mJ_{\mu\nu} \biggr) 
\nonumber \\
& &\quad \times {1\over{\square+m^{2}}} 
\biggl( {\delta^{@,@,@,[@,@,\mu}}_{\rho} {\delta^{\nu@,@,]}}_{\sigma}
+\frac{k-1}{\square+km^{2}}@,@,@, 
\partial^{@,@,[@,@,\mu}\partial_{@,@,@,[@,@,\rho} 
{\delta^{\nu@,@,]}}_{\sigma@,]} \biggr) 
\nonumber \\
& &\quad \times 
\biggl( \Bigl(m-\frac{u}{k}\Bigr) F^{\rho\sigma}+mJ^{\rho\sigma} \biggr) 
\nonumber \\
& & \biggl. \biggl.
+{1\over2k}(\partial_{\mu}\varphi+uA_{\mu})
 (\partial^{\mu}\varphi+uA^{\mu})+{\cal L}_{b}+A_{\mu}@,@,j^{\mu} 
\biggr\} \biggr]  \,
\footnotemark[3] 
\nonumber \\
&=& N_{3} \int{\frak D}@!@!A_{\mu}@,@,@,@,@, {\frak D}\varphi@,@,@,@,@,@,@, 
{\frak D}b 
@,@, \exp \biggl[ \, i \int d^{4}x \biggl\{ 
-{1\over4}F_{\mu\nu}F^{\mu\nu} 
\biggr. \biggr.
\nonumber \\
& &
+{1\over4}\Bigl(m-\frac{u}{k}\Bigr)^{2}
F_{\mu\nu}@,{k\over{\square+km^{2}}}@,F^{\mu\nu} 
\nonumber \\
& &
+{1\over2k}(\partial_{\mu}\varphi+uA_{\mu})
 (\partial^{\mu}\varphi+uA^{\mu})
+{\cal L}_{b}
\nonumber \\
& &
+A_{\mu}@,@,j^{\mu} 
-\Bigl(m-\frac{u}{k}\Bigr)A_{\mu}@, 
{km\over{\square+km^{2}}}@,@,@,@,@, j^{\mu}
\nonumber \\
& & \biggl. \biggl. -{1\over2}@,@,@,@,@,@,@,j_{\mu}@, 
\frac{(k-1)m^{2}}{(\square+km^{2})(\square+m^{2})}@,@,@,@,@, j^{\mu}
%\nonumber \\
%& &
+{1\over4}J_{\mu\nu}@, \frac{m^{2}}{\square+m^{2}}@, J^{\mu\nu}
\biggr\} \biggr] \,.
\label{22}
\end{eqnarray}
%(22)
%
%
\footnotetext[3]{The brackets $[\;\;\:]$ that enclose two indices stand for 
the antisymmetrization defined by 
$\,X^{[@,\mu}Y^{\nu]}\equiv X^{\mu}Y^{\nu}-X^{\nu}Y^{\mu}$.}Here, 
Eqs. (\ref{4}) and (\ref{5}) have been used to obtain the last form. 
As will be seen below, the generating functional $Z$ is 
actually independent of the parameters $k$ and $u$ as well as the gauge 
parameter $\alpha$. 
Taking into account this fact, 
we now set $k=1$ and $u=m$ so that all the non-local terms of $A_{\mu}$ and 
$j^{\mu}$ in Eq. (\ref{22}) vanish. 
Then, the integrand in the exponent of Eq. (\ref{22}) 
reduces to the sum of the non-local term 
$\dfrac{1}{4} m^{2}J_{\mu\nu}({\square+m^{2}})^{-1}J^{\mu\nu}$  
and the Lagrangian (with gauge-fixing and current terms) that defines 
the (quantum) Abelian Stueckelberg formalism \cite{Stu}. 
This Lagrangian clearly describes a vector field that has mass $m$ and 
couples with the current $j^{\mu}$. 
It should be noted that the MAATGT with ATC is not equivalent to 
the Abelian Stueckelberg formalism, since the MAATGT with ATC involves 
the non-local term of $J^{\mu\nu}$ in addition to the Lagrangian of 
the Abelian Stueckelberg formalism.

The integration over $\varphi$ in Eq. (\ref{22}) can be carried out without 
any difficulty. In the case $\alpha\neq0@,@,$, the integral over 
$b$ in  Eq. (\ref{22}) becomes a simple Gaussian integral. 
After the integrations over $\varphi$ and $b$, 
we can perform the integration over $A_{\mu}$ to obtain a final form of $Z$. 
In the case $\alpha=0@,@,$, the integration over $b$ gives the delta function 
$\prod_{x}\delta(\partial^{\mu}A_{\mu})$, which allows us to remove 
the terms proportional to 
$\partial^{\mu}A_{\mu}$ from the exponent of Eq. (\ref{22}).
Removing them, we express $\prod_{x}\delta(\partial^{\mu}A_{\mu})$ 
in the form of the integral over $b$ again. 
After the integrations over $\varphi$ and 
$A_{\mu}@,@,$, we carry out the integration over $b$ to get a final form
of $Z@,$. 
In the both cases $\alpha\neq0$ and $\alpha=0@,@,$, the resulting expression 
of $Z$ takes the following form:  
\begin{eqnarray}
Z[J^{\mu\nu}, j^{\mu}@,]&=& 
N_{4} 
\exp \biggl[ \, i \int d^{4}x \biggl\{ 
-{1\over2}@,@,@,@,@,@,@,
j_{\mu}@, \frac{1}{\square+m^{2}}@,@,@,@,@, j^{\mu}
+{1\over4}J_{\mu\nu}@, \frac{m^{2}}{\square+m^{2}}@, J^{\mu\nu}
\biggr\} \biggr] \,, 
\label{23}
\end{eqnarray}
%(23)
%
from which we see that the generating functional $Z$ is indeed independent of 
the parameters $\alpha$, $k(\neq0)$ and $u@,$.  
In the process of deriving Eq. (\ref{23}), we understand that the first 
integrand in the exponent of Eq. (\ref{23}) originates in propagation of 
$A_{\mu}@,@,$, while the second integrand originates in propagation of 
$B_{\mu\nu}@,@,$.

Next we suppose $k=0@,@,$. In this case, the integration over 
$\beta_{\mu}$ in Eq. (\ref{20}) yields the delta function 
$\prod_{x,@,@,\nu}\delta@,(\partial^{\mu}B_{\mu\nu}+\partial_{\nu}\varphi
+uA_{\nu})$, which makes it possible to replace 
the term $\partial^{\mu}B_{\mu\nu}\partial_{\rho}B^{\rho\nu}$ 
in the exponent of Eq. (\ref{20}) by 
$B^{\rho\nu}\partial_{\rho}(\partial_{\nu}\varphi+uA_{\nu})
={1\over2}uB^{\rho\nu}F_{\rho\nu}@,@,$. After this replacement, the 
integration over $B_{\mu\nu}$ in Eq. (\ref{20}) can be done, leading to 
\begin{eqnarray}
Z[J^{\mu\nu}, j^{\mu}@,]&=& 
N_{5} \int{\frak D}@!@!A_{\mu}@,@,@,@,@, {\frak D}\beta_{\mu} @,@,@,@,@, 
{\frak D}\varphi @,@,@,@,@,@,@, {\frak D}b 
@,@, \exp \biggl[ \, i \int d^{4}x \biggl\{ 
-{1\over4}F_{\mu\nu}F^{\mu\nu} 
\biggr. \biggr.
\nonumber \\
& & 
+{1\over4} \biggl( \Bigl(m-\frac{u}{2}\Bigr) F_{\mu\nu}
+\partial_{@,@,@,[@,@,\mu} \beta_{\nu@,@,]} +mJ_{\mu\nu} \biggr) 
\nonumber \\
& &\quad \times {1\over{\square+m^{2}}} 
\biggl( \Bigl(m-\frac{u}{2}\Bigr) F^{\mu\nu}
+\partial^{@,@,[@,@,\mu} \beta^{\nu@,@,]} +mJ^{\mu\nu} \biggr) 
\nonumber \\
& & -\beta^{\mu}(\partial_{\mu}\varphi+uA_{\mu})
 +{\cal L}_{b}+A_{\mu}@,@,j^{\mu} 
\biggr\} \biggr] \,. 
\label{24}
\end{eqnarray}
%(24)
%
Since the integration over $\varphi$ in Eq. (\ref{24}) yields the delta 
function $\prod_{x}\delta@,(\partial^{\mu}\beta_{\mu})$, the term 
$-{1\over2}\partial_{\mu}\beta_{\nu}(\square+m^{2})^{-1}\partial^{\nu}
\beta^{\mu}$ in the exponent of Eq. (\ref{24}) can be removed from it. 
Then, the integration over $\beta_{\mu}$ reduces to a Gaussian integration 
and is carried out to obtain 
\begin{eqnarray}
Z[J^{\mu\nu}, j^{\mu}@,]&=& 
N_{6} \int{\frak D}@!@! A_{\mu}@,@,@,@,@, {\frak D}\varphi@,@,@,@,@,@,@,
{\frak D}b 
@,@, \exp \biggl[ \, i \int d^{4}x \biggl\{ 
-{1\over4}F_{\mu\nu}F^{\mu\nu} 
\biggr. \biggr.
\nonumber \\
& &
-{u\over2}\Bigl(m-\frac{u}{2}\Bigr)
F_{\mu\nu}@, \square^{-1} @,F^{\mu\nu} 
\nonumber \\
& &
+{1\over2}(\partial_{\mu}\varphi+uA_{\mu})
\big(1+m^{2}\square^{-1} \big)
 (\partial^{\mu}\varphi+uA^{\mu})
+{\cal L}_{b}
\nonumber \\
& &
+A_{\mu} @,@,@, j^{\mu} 
+umA_{\mu}@, \square^{-1}  j^{\mu}
\nonumber \\
& & \biggl. \biggl. 
+{1\over2}@,@,@,@,@,@,@,j_{\mu}@, 
\frac{m^{2}}{\square@,(\square+m^{2})}@,@,@,@,@, j^{\mu}
+{1\over4}J_{\mu\nu}@, \frac{m^{2}}{\square+m^{2}}@, J^{\mu\nu}
\biggr\} \biggr] \,.
\label{25}
\end{eqnarray}
%(25)
%
Here, Eqs. (4) and (5) have been used. 
The integrations over $\varphi$, $b$ and $A_{\mu}$ in Eq. (\ref{25}) 
can be carried out in manners similar to those in Eq. (\ref{22}) and 
we arrive at Eq. (\ref{23}) again. Hence, 
in the both cases $k\neq0$ and $k=0@,@,$,  the generating 
functional $Z$ turns out to be the form of Eq. (\ref{23}).

Now, introducing a space-like constant vector $n^{\mu}$, we formally 
solve Eq. (\ref{4}) for $J^{\mu\nu}$ in terms of $j^{\mu}$ and 
$n^{\mu}@,@,@,$: 
\begin{eqnarray}
J^{\mu\nu}={1\over{n@!@!@!@!@!\cdot@!@!@!@!@!\partial}}@,@, 
(n^{\mu}j^{\nu}-n^{\nu}j^{\mu}) \,,\label{26}
\end{eqnarray}
%(26)
%
where $n@!@!@!@!\cdot@!@!@!@!\partial\equiv n^{\mu}\partial_{\mu}@,@,$. 
The constant vector 
$n^{\mu}$ is nothing but a set of integration constants in solving 
Eq. (\ref{4}). It is easy to check, with the conservation law in 
Eq. (\ref{5}), that Eq. (\ref{26}) is indeed a solution of Eq. (\ref{4}). 
Substituting Eq. (\ref{26}) into Eq. (\ref{23}), we have 
\begin{eqnarray}
Z[\,@! j^{\mu}@,]&=& 
N_{4} 
\exp \biggl[ \, i \int d^{4}x \biggl\{ 
-{1\over2}@,@,@,@,@,@,@,j_{\mu} \biggl(
\frac{1}{\square+m^{2}}@,@, {\delta^{\mu}}_{\nu} 
\biggr. \biggr.
\nonumber \\ 
& &
\biggl. \biggl. 
+\frac{m^{2}}{\square+m^{2}} 
\frac{n^{2}}{(n@!@!@!@!@!\cdot@!@!@!@!@!\partial)^{2}}
\Bigl({\delta^{\mu}}_{\nu} -\frac{n^{\mu}n_{\nu}}{n^{2}} \Bigr)
\biggr)@,@,j^{\nu} \biggr\} \biggr] \,, 
\label{27}
\end{eqnarray}
%(27)
%
where $n^{2}\equiv n_{\mu}n^{\mu}$. This $Z$ agrees with 
the generating functional obtained by Suzuki \cite{Suz} in his study of 
a low energy effective theory of 
SU(2) QCD, in which the constant vector $n^{\mu}$ 
was introduced in accordance with Zwanziger's formulation for electric 
and magnetic charges \cite{Zwan}.

In order to see what kind of static potential is derived from the effective 
action $W[\,@! j^{\mu}@,]=-i\ln (Z[\,@! j^{\mu}@,]/Z[@,@,0@,@,]@,)$  
with Eq. (\ref{27}), we take $j^{\mu}$ as  
the static current 
$j^{\mu}_{Q}(x)\equiv{\delta^{\mu}}_{0} Q \left\{ \delta^{3}(\boldsymbol{x}
-\boldsymbol{r})-\delta^{3}(\boldsymbol{x}) \right\}$ satisfying 
the conservation law (\ref{5}). Here $Q$ and $-Q$ are point charges at 
$\boldsymbol{x}=\boldsymbol{r}$ and $\boldsymbol{x}=\boldsymbol{0}$, 
respectively. 
Substituting $j^{\mu}_{Q}$ into $W[\,@! j^{\mu}@,]$ 
and choosing a constant vector $(0,\, \boldsymbol{n})$ with the condition 
$\boldsymbol{n}/\!/\boldsymbol{r}$ \footnotemark[4] to be $n^{\mu}$,  
we can calculate 
the effective potential, $V_{\rm eff}$, defined by 
$-V_{\rm eff}\int dx^{0} =W[\,@! j^{\mu}_{Q}@,]$. 
As a result of the calculation, we find that the part of  
$W[\,@! j^{\mu}_{Q}@,]$ 
that is independent of $n^{\mu}$ leads to the Yukawa potential, 
while the part of $W[\,@! j^{\mu}_{Q}@,]$ that is dependent on 
$n^{\mu}$ leads to the linearly rising potential.  
This shows that propagation of $A_{0}$ gives rise to the Yukawa potential, 
while propagation of $B_{0i}$ $(@,@,@,i=1,2,3 @,@,@,)$ gives rise to 
the linear potential. 
The precise form of the effective potential $V_{\rm eff}$ is \cite{Suz,SST}
\footnotetext[4]{The parallel condition $\boldsymbol{n}/\!/\boldsymbol{r}$ 
is set by taking into account the axial symmetry of the system and 
the minimum energy condition \cite{SST}.}
\begin{eqnarray}
V_{\rm eff}(r)=-\frac{Q^{2}}{4\pi} \frac{e^{-mr}}{r}
+\frac{Q^{2}m^{2}}{8\pi} 
\biggr[ \ln \biggl(1+\frac{\Lambda^{2}}{m^{2}} \biggr) \biggr] r 
+\mbox{\rm constant}\,. 
\label{28}
\end{eqnarray}
%(28)
%
Here $r\equiv|\boldsymbol{r}|@,$, and $\Lambda$ is an ultraviolet cut-off. 
Therefore, the MAATGT with ATC yields the composite of the 
Yukawa and the linear potentials, describing confinement of 
the point charges.  
We can also derive an effective potential similar to Eq. (\ref{28}) from 
Eq. (\ref{23}) by means of an alternative method in which  
$J^{\mu\nu}$ is taken as the vorticity tensor 
current defined in terms of string variables. 
This method will be mentioned in the last section.

\section{Noncovariant quantization in axial gauge}

Let us turn to the noncovariant quantization of $B_{\mu\nu}$ in axial gauge. 
We now utilize $n^{\mu}$ introduced in Eq. (\ref{26}) as a space-like 
constant vector (or an axis) characterizing the axial gauge \cite{BEMSSZ} 
and choose the noncovariant gauge-fixing term 
\begin{align}
{\widehat{\cal L}}_{\rm G1} &=
-i\boldsymbol{\delta} \left[ B_{\mu\nu}n^{\mu}\bar{\rho}^{\nu} \right] 
\nonumber \\
&=B_{\mu\nu}n^{\mu}\beta^{\nu}
-{i\over2}(\partial_{\mu}\rho_{\nu}-\partial_{\nu}\rho_{\mu})
(n^{\mu}\bar{\rho}^{\nu}-n^{\nu}\bar{\rho}^{\mu}) \,,  
\label{29}
\end{align}
%(29)
%
instead of the covariant gauge-fixing term in Eq. (\ref{13}). 
(The BRST transformation $\boldsymbol{\delta}$ in Eq. (\ref{29}) is 
the one defined in Sec. 3.)
Here the dimension of $n^{\mu}$ can be determined to be either dimensionless 
or mass-dimension one, depending on the dimensions of the fields 
$\bar{\rho}_{\mu}$ and $\beta_{\mu}@,@,$. 
The gauge-fixing term ${\widehat{\cal L}}_{\rm G1}$, 
which realizes a noncovariant analogue of the Landau gauge 
condition, i.e. $n^{\mu}B_{\mu\nu}=0\,\footnotemark[1]$\!,  
remains invariant under the secondary 
gauge transformation $\delta\rho_{\mu}=\partial_{\mu}\varepsilon @,@,, \;
\delta\bar{\rho}_{\mu}=n_{\mu}\bar{\varepsilon}@,@,@,$. 
\footnotetext[1]{ We may, of course, choose 
the axial gauge condition $n^{\mu}B_{\mu\nu}+{1\over2}k\beta_{\nu}=0$ 
with a non-zero gauge parameter $k$.  
However, the Landau-type gauge condition $n^{\mu}B_{\mu\nu}=0$ is essential 
for the following discussion and so we consider only it.}Consequently, 
to quantize $\rho_{\mu}$ and $\bar{\rho}_{\mu}@,@,$, 
we have to introduce a gauge-fixing term that breaks 
the invariance of ${\widehat{\cal L}}_{\rm G1}$ under 
this gauge transformation. 
As a suitable gauge-fixing term, we now adopt the gauge-fixing term 
in Eq. (\ref{14}) 
again. In addition, we use the gauge-fixing term in Eq. (\ref{15}) to complete 
the gauge-fixing. (Thereby $A_{\mu}$ is quantized in the covariant 
gauge.) The complete gauge-fixing term is thus 
\begin{eqnarray}
& & \widehat{\cal L}_{\rm G1}+{\cal L}_{\rm G2}+{\cal L}_{\rm G3} 
\nonumber \\
&=& 
-\beta^{\nu}(-n^{\mu}B_{\mu\nu}+\partial_{\nu}\varphi+uA_{\nu})
\nonumber \\
& & 
-i\bar{\rho}^{\nu} \{(-n@!@!@!@!@!\cdot@!@!@!@!@!\partial+um)\rho_{\nu}
+n^{\mu}\partial_{\nu}\rho_{\mu}
+\partial_{\nu}\chi 
+u\partial_{\nu}c \}
\nonumber \\
& &
-i\rho^{\nu}(\partial_{\nu}\bar{\chi}-m\partial_{\nu} \bar{c})
-\bar{\sigma}@, \square @,
\sigma +i\bar{c}@,@,@,@, \square @, c 
+{\cal L}_{b} 
\nonumber \\
& & 
+\,\mbox{total derivative}\,,
\label{30}
\end{eqnarray} 
%(30)
%
where the dimension of the parameter $u$ is determined to be the same as 
that of $n^{\mu}@,$.

The generating functional we are now concerned with is 
\begin{eqnarray}
{\widehat Z}[J^{\mu\nu}, j^{\mu}@,] 
= \widehat{N}_{0} \int {\frak D}@!@!{\cal M} 
\exp\left[\,i\int d^{4}x ({\cal L}_{0}
+\widehat{\cal L}_{\rm G1}+{\cal L}_{\rm G2}+{\cal L}_{\rm G3} ) \right] . 
\label{31}
\end{eqnarray}
%(31)
%
Here and hereafter, $\widehat{N}_{i}$ $(@,@,@,i=0,1,2,3,4 @,@,@,)$ denote  
constants. 
Along a line similar to obtaining Eq. (\ref{20}) from Eq. (\ref{18}), 
we carry out the integrations over the ghost fields $\rho_{\mu}@,@,$, 
$\bar{\rho}_{\mu}@,@,$, $\chi$ and $\bar{\chi}$ in the following way:  
First, note that the delta functions 
$\prod_{x}\delta(\partial^{\nu}\bar{\rho}_{\nu})$ and 
$\prod_{x}\delta(\partial^{\nu}\rho_{\nu})$ given by 
the integrations over $\chi$ and $\bar{\chi}$ in Eq. (\ref{31})
enable us to remove the three terms 
$-i\bar{\rho}^{\nu}n^{\mu}\partial_{\nu}\rho_{\mu}@,@,$, 
$-iu\bar{\rho}^{\nu}\partial_{\nu}c$  
and $im\rho^{\nu}\partial_{\nu}\bar{c}$ from the exponent of Eq. (\ref{31}). 
Being removed them, the integration over $\rho_{\mu}$ and 
$\bar{\rho}_{\mu}$ yields 
$\{ \det(-n@!@!@!@!@!\cdot@!@!@!@!@!\partial+um) \}^{4} 
\exp \! \left[\, i\int d^{4}x \, \bar{\chi} 
\{ i @,@,@, \square@,@,(-n@!@!@!@!@!\cdot@!@!@!@!@!\partial+um)^{-1} \} 
\chi \right]$, and 
the integration over $\chi$ and $\bar{\chi}$ gives 
$\det\{ \square@,@,(-n@!@!@!@!@!\cdot@!@!@!@!@!\partial+um)^{-1} \}@,@,$. 
After the integrations over 
$\sigma@,$, $\bar{\sigma}@,$, $c@,$ and $\bar{c}@,@,@,$, 
the generating functional $\widehat{Z}$ becomes 
\begin{eqnarray}
{\widehat Z}[J^{\mu\nu}, j^{\mu}@,]&=& 
{\widehat N}_{1} \int{\frak D}@!@!B_{\mu\nu}@,@,@,@,@, 
{\frak D}@!@!A_{\mu}@,@,@,@,@,
{\frak D}\beta_{\mu}@,@,@,@,@, {\frak D}\varphi @,@,@,@,@,@,@, {\frak D}b 
\nonumber \\
& & \times \exp \biggl[ \, i \int d^{4}x \biggl\{ 
-{1\over4}B_{\mu\nu}(\square+m^{2}) B^{\mu\nu}
\biggr. \biggr.
\nonumber \\
& & 
-{1\over2}\partial^{\mu}B_{\mu\nu}\partial_{\rho}B^{\rho\nu} 
+\frac{m}{2} B_{\mu\nu}F^{\mu\nu} -{1\over4}F_{\mu\nu}F^{\mu\nu} 
\nonumber \\
& & +\beta^{\nu}(n^{\mu}B_{\mu\nu}-\partial_{\nu}\varphi
-uA_{\nu}) +{\cal L}_{b}
\nonumber \\
& &
\biggl. \Bigl.
+{1\over2}mB_{\mu\nu}J^{\mu\nu}+A_{\mu}@,@,j^{\mu} 
\biggr\} \biggr] \;.
\label{32}
\end{eqnarray}
%(32)
%

We now assume that the current $J^{\mu\nu}$ takes the form of Eq. (\ref{26}). 
Then, we see that 
$B_{\mu\nu}J^{\mu\nu}=-2j^{\nu}(n@!@!@!@!@!\cdot@!@!@!@!@!\partial)^{-1}
(n^{\mu}B_{\mu\nu})$ 
up to a total derivative. Substituting this equation into 
Eq. (\ref{32}) and noting the fact that the integration over $\beta_{\mu}$ in 
Eq. (\ref{32}) yields the delta function 
$\prod_{x,@,@,\nu}\delta@,(n^{\mu}B_{\mu\nu}-\partial_{\nu}\varphi
-uA_{\nu})$, we can write Eq. (\ref{32}) as 
\begin{eqnarray}
{\widehat Z}[\,@! j^{\mu}@,]&=& 
{\widehat N}_{1} \int{\frak D}@!B_{\mu\nu}@,@,@,@, {\frak D}@!A_{\mu}@,@,@,@, 
{\frak D}\beta_{\mu} @,@,@,@, {\frak D}\varphi @,@,@,@,@,@, {\frak D}b 
\nonumber \\
& & \times \exp \biggl[ \, i \int d^{4}x \biggl\{ 
-{1\over4}B_{\mu\nu}(\square+m^{2}) B^{\mu\nu}
\biggr. \biggr.
\nonumber \\
& & 
-{1\over2}\partial^{\mu}B_{\mu\nu}\partial_{\rho}B^{\rho\nu} 
+\frac{m}{2} B_{\mu\nu}F^{\mu\nu} -{1\over4}F_{\mu\nu}F^{\mu\nu} 
\nonumber \\
& & +\beta^{\nu}(n^{\mu}B_{\mu\nu}-\partial_{\nu}\varphi
-uA_{\nu}) +{\cal L}_{b}
\nonumber \\
& &
\biggl. \biggl.
+@,@,@,j^{\mu} \big( 1-um(n@!@!@!@!@!\cdot@!@!@!@!@!\partial)^{-1} \big) 
A_{\mu} \biggr\} \biggr] \;,
\label{33}
\end{eqnarray}
%(33)
%
with the aid of Eq. (\ref{5}).  
It should be emphasized here that choosing $n^{\mu}$ in Eq. (\ref{26}) 
to be the axis of the axial gauge condition 
has made it possible to replace ${1\over2}mB_{\mu\nu}J^{\mu\nu}$ in 
the exponent of  Eq. (\ref{32}) by 
$-umj^{\mu}(n@!@!@!@!@!\cdot@!@!@!@!@!\partial)^{-1}A_{\mu}@,@,$. 
As a result of the replacement, the tensor current term  
disappeared on the surface from $\widehat{Z}$.

Next we would like to carry out the integration over $B_{\mu\nu}@,@,$. 
Unlike the case $k=0$ in the covariant quantization, it is impossible 
to replace the term $\partial^{\mu}B_{\mu\nu}\partial_{\rho}B^{\rho\nu}$ 
in the exponent of Eq. (\ref{33}) by a suitable term. 
This is due to the fact that the delta function 
$\prod_{x,@,@,\nu}\delta@,(n^{\mu}B_{\mu\nu}-\partial_{\nu}\varphi
-uA_{\nu})$ that occurs in Eq. (\ref{33}) by the integration over 
$\beta_{\mu}$ is useless to the replacement. 
Thus, instead of such an attempt, 
we perform the integrations over $B_{\mu\nu}@,@,$, $\beta_{\mu}$ and 
$\varphi$ simultaneously by noting the fact that these 
integrations can be treated as a {\it single} Gaussian integration over 
the set of $B_{\mu\nu}@,@,$, $\beta_{\mu}$ and $\varphi$. 
The calculation is slightly tedious, but can be done straightforwardly 
by making reference to the derivation  
of free propagators in quantum electrodynamics in 
the axial gauge \cite{BEMSSZ}. The result reads 
\begin{eqnarray}
{\widehat Z}[\,@! j^{\mu}@,]&=& 
\widehat{N}_{2} \int {\frak D}@!A_{\mu}@,@,@,@, {\frak D}b 
@,@, \exp \biggl[ \, i \int d^{4}x \biggl\{ 
-{1\over4}F_{\mu\nu}F^{\mu\nu} 
\nonumber \\
& & 
+{m^{2}\over2} F_{\mu\nu}{\varDelta^{\mu\nu}}_{\rho\sigma} 
F^{\rho\sigma}-2um F_{\mu\nu}{\varDelta^{\mu\nu}}_{\rho} A^{\rho}
\nonumber \\
& &
+2u^{2} A_{\mu}{\varDelta^{\mu}}_{\nu} A^{\nu} 
+{\cal L}_{b}
\biggl. \biggl.
@,@,@,+ 
@,@,@,@,j^{\mu} \big( 1-um(n@!@!@!@!@!\cdot@!@!@!@!@!\partial)^{-1} \big) 
A_{\mu} \biggr\} \biggr] \;,
\label{34}
\end{eqnarray}
%(34)
%
with 
\begin{eqnarray}
{\varDelta^{\mu\nu}}_{\rho\sigma} &\equiv& 
\frac{1}{4(\square+m^{2})} \Big[ 
{\delta^{\mu}}_{[@,@,\rho} {\delta^{\nu}}_{\sigma@,]} 
\nonumber \\ 
& & -\frac{1}{(n@!@!@!@!@!\cdot@!@!@!@!@!\partial)^{2}+m^{2}n^{2}}
\Big\{ (n@!@!@!@!@!\cdot@!@!@!@!@!\partial)@, 
{@,@,@,\delta^{[@,@,\mu}}_{[@,@,\rho} 
\big( n^{\nu@,@,]@,}\partial_{\sigma@,]}
+n_{\sigma@,]}\partial^{\nu@,@,]}@, \big)
\nonumber \\ 
& & 
-n^{2}{@,@,@,\delta^{[@,@,\mu}}_{[@,@,\rho} @,\partial^{\nu@,@,]} 
\partial_{\sigma@,]} 
+m^{2}{@,@,@,\delta^{[@,@,\mu}}_{[@,@,\rho} @,n^{\nu@,@,]} n_{\sigma@,]} 
+n^{[@,@,\mu}@,n_{[@,@,\rho} @,\partial^{\nu@,@,]} \partial_{\sigma@,]} 
\Big\} \Big]\,,
\label{35}
\\ 
& &
\nonumber \\
{\varDelta^{\mu\nu}}_{\rho} &\equiv& 
\frac{1}{4 \big( (n@!@!@!@!@!\cdot@!@!@!@!@!\partial)^{2}+m^{2}n^{2} \big) } 
\Big[ (n@!@!@!@!@!\cdot@!@!@!@!@!\partial)@, 
{\delta^{@,@,@,[@,@,\mu}}_{\rho}@, 
\partial^{\nu@,@,]} 
+m^{2} {\delta^{@,@,@,[@,@,\mu}}_{\rho}@, n^{\nu@,@,]} 
\nonumber \\
& & 
+m^{2}(n@!@!@!@!@!\cdot@!@!@!@!@!\partial)^{-1} n_{\rho}@, 
n^{[@,@,\mu}\partial^{\nu@,@,]} 
\Big] \,,
\label{36}
\\
& & 
\nonumber \\
{\varDelta^{\mu}}_{\nu} &\equiv& 
-\frac{m^{2}}{4 \big( (n@!@!@!@!@!\cdot@!@!@!@!@!\partial)^{2}
+m^{2}n^{2} \big) } 
\Big[ (\square+m^{2}){\delta^{\mu}}_{\nu} -\partial^{\mu}\partial_{\nu}
\nonumber \\ 
& &
-m^{2}(n@!@!@!@!@!\cdot@!@!@!@!@!\partial)^{-1}
(n^{\mu}@,\partial_{\nu}+n_{\nu}@,\partial^{\mu})
+m^{2}  n^{\mu}n_{\nu} \square (n@!@!@!@!@!\cdot@!@!@!@!@!\partial)^{-2}
\Big] \,.
\label{37}
\end{eqnarray}
%(35)(36)(37)
%
Since the integrations over $B_{\mu\nu}@,@,$, $\beta_{\mu}$ and $\varphi$ 
never spoil the symmetry under the gauge transformation 
$\delta A_{\mu}=\partial_{\mu}\lambda@,@,$, the integrand in the exponent of 
Eq. (\ref{34}), apart from the gauge-fixing term ${\cal L}_{b}@,@,$, must be 
invariant under this transformation up to total derivatives; 
in fact, Eq. (\ref{34}) can be expressed in the form 
\begin{eqnarray}
{\widehat Z}[\,@! j^{\mu}@,]&=& 
\widehat{N}_{2} \int {\frak D}@!A_{\mu}@,@,@,@, {\frak D}b 
@,@, \exp \biggl[ \, i \int d^{4}x 
\nonumber \\
& &
\times \biggl\{ 
-{1\over4}F_{\mu\nu}
\big( 1-2um(n@!@!@!@!@!\cdot@!@!@!@!@!\partial)^{-1}
-u^{2}m^{2}(n@!@!@!@!@!\cdot@!@!@!@!@!\partial)^{-2} \big)
{\varXi^{\mu}}_{\rho} F^{\rho\nu} 
\nonumber \\
& &
+ {\cal L}_{b}
+@,@,j^{\mu} \big( 1-um(n@!@!@!@!@!\cdot@!@!@!@!@!\partial)^{-1} \big) A_{\mu} 
\biggr\} \biggr] \;,
\label{38}
\end{eqnarray}
%(38)
%
with 
\begin{eqnarray}
{\varXi^{\mu}}_{\rho} \equiv
{\delta^{\mu}}_{\rho}
+\frac{m^{2}}{(n@!@!@!@!@!\cdot@!@!@!@!@!\partial)^{2}+m^{2}n^{2}} 
\big( 2n^{\mu}n_{\rho}-{\delta^{\mu}}_{\rho} n^{2} \big) \,.
\label{39}
\end{eqnarray}
%(39)
%
It is easy to show 
$\int d^{4}x F_{\mu\nu}(n@!@!@!@!@!\cdot@!@!@!@!@!\partial)^{-1}
{\varXi^{\mu}}_{\rho} F^{\mu\nu}=0$ by partial integrations done  
in the left-hand side of this equation. Using this formula and changing  
integration variables of Eq. (\ref{38}) from $A_{\mu}$ to 
$\big( 1-um(n@!@!@!@!@!\cdot@!@!@!@!@!\partial)^{-1} \big) A_{\mu}@,@,@,$, 
we now rewrite Eq. (\ref{38}) as 
\begin{eqnarray}
{\widehat Z}[\,@! j^{\mu}@,]&=& 
\widehat{N}_{3} \int {\frak D}@!A_{\mu}@,@,@,@, {\frak D}b 
@,@, \exp \biggl[ \, i \int d^{4}x 
\biggl\{ 
-{1\over4}F_{\mu\nu} {\varXi^{\mu}}_{\rho} F^{\rho\nu} 
%\nonumber \\
%& &
+{\cal L}_{b}+ A_{\mu}@,@,j^{\mu} 
\biggr\} \biggr] \;.
\label{40}
\end{eqnarray}
%(40)
%
Notice here that the parameter $u$ disappeared from ${\widehat Z}$.  
This shows that the generating functional $\widehat{Z}$ is actually 
independent of $u$, 
as might be expected from a similar result found in Sec. 3.

In the case $\alpha\neq0@,@,$, Eq. (\ref{40}) can be written as 
\begin{eqnarray}
{\widehat Z}[\,@! j^{\mu}@,]&=& 
\widehat{N}_{4} \int {\frak D}@!A_{\mu}@,@, \exp \biggl[ \, i \int d^{4}x 
\biggl\{ \,{1\over2} A_{\mu} 
\bigg( \square {\delta^{\mu}}_{\nu}-\bigg( 1-{1\over\alpha} \bigg)
\partial^{\mu} \partial_{\nu} 
\bigg. \biggr. \biggr. 
\nonumber \\ 
& & 
-\frac{m^{2}}{(n@!@!@!@!@!\cdot@!@!@!@!@!\partial)^{2}+m^{2}n^{2}} 
\Big\{ \big( n^{2}\square-(n@!@!@!@!@!\cdot@!@!@!@!@!\partial)^{2}\big)
{\delta^{\mu}}_{\nu}
-n^{2}@,\partial^{\mu}\partial_{\nu}
\nonumber \\
& & 
\biggl. \biggl. \bigg.  
-n^{\mu}@,n_{\nu}@,\square 
+(n@!@!@!@!@!\cdot@!@!@!@!@!\partial)(n^{\mu}@,\partial_{\nu}
+n_{\nu}@,\partial^{\mu})
\Big\} \bigg) A^{\nu}
+A_{\mu}@,@,j^{\mu} 
\biggr\} \biggr] \;,
\label{41}
\end{eqnarray}
%(41)
%
after the integration over $b$.  
The non-local Lagrangian for $A_{\mu}$ in the exponent of Eq. (\ref{41}) 
has already been found by Suzuki \cite{Suz} through Zwanziger's formulation. 
It should be stressed that in our discussion,  
the non-local Lagrangian for $A_{\mu}$ has been obtained by virtue of 
the appropriate choice of the axis in the axial gauge condition. 
Carrying out the integration over $A_{\mu}$ in Eq. (\ref{41}), 
we arrive at a final form of $\widehat{Z}$.  
In the case $\alpha=0@,@,$, we first perform the integration over $A_{\mu}$ 
by taking into account the fact that the integration over $b$ gives 
the delta function $\prod_{x}\delta(\partial^{\mu}A_{\mu})$.  
After that, the integration over $b$ leads to 
a final form of $\widehat{Z}$. 
In the both cases $\alpha\neq0$ and $\alpha=0@,@,$, the resulting 
expression of $\widehat{Z}$ is exactly same as $Z$ in 
Eq. (\ref{27}), apart from a constant such as $N_{4}$. 
Therefore, as long as $J^{\mu\nu}$ takes the specific form of 
Eq. (\ref{26}), 
the generating functionals $Z$ and ${\widehat Z}$ turn out to be the same  
up to their overall constants, i.e.   
$Z[\,@! j^{\mu}@,]/Z[@,@,0@,@,]
={\widehat Z}[\,@! j^{\mu}@,]/{\widehat Z}[@,@,0@,@,]@,@,$.

\section{Summary and Discussion}

In this letter, we have discussed a quantum-theoretical aspect of the MAATGT 
with ATC consisting of an antisymmetric tensor field $B_{\mu\nu}$, a vector  
field $A_{\mu}$, and their associated currents $J^{\mu\nu}$ and $j^{\mu}$. 
The quantization of $B_{\mu\nu}$ has been performed both in the covariant 
gauge with a gauge parameter and in the axial gauge of the Landau type, 
while the quantization of $A_{\mu}$ has been performed only in 
the covariant gauge with a gauge parameter.  
In the covariant quantization of $B_{\mu\nu}$, 
we have derived a generating functional written in terms of 
the currents $J^{\mu\nu}$ and $j^{\mu}$ (see Eq. (\ref{23})) 
by carrying out the path-integrations over all the relevant fields, 
including the ghost and auxiliary fields. 
Then, taking $J^{\mu\nu}$ to be the specific form of Eq. (\ref{26}), 
we have obtained the generating functional written in terms of $j^{\mu}$ 
alone (see Eq. (\ref{27})), 
which agrees with that found earlier by using Zwanziger's formulation 
\cite{Zwan}.  
This generating functional actually leads to the composite of 
the Yukawa and the linear potentials.
In the noncovariant quantization of $B_{\mu\nu}$ in the axial gauge, 
we were able to directly find a non-local Lagrangian for $A_{\mu}$ 
with the only current term $A_{\mu}@,@,j^{\mu}$ (see Eq. (\ref{41})) 
by carrying out the path-integrations over all the relevant fields except 
$A_{\mu}$.  
This is an advantage of taking the axial gauge with a suitable axis. 
It has also been seen that the generating functional with this non-local 
Lagrangian reduces to the generating functional in Eq. (\ref{27}) up to an 
overall constant; consequently, the quantization of $B_{\mu\nu}$ in 
the covariant gauge and that in the axial gauge yield the same 
physical result.

The generating functional in Eq. (\ref{23}) is essentially the same as  
Antonov and Ebert have obtained in Ref. \cite{AE}. 
They derived it in a complicated manner by taking the unitary gauge for 
$B_{\mu\nu}$, in which the vector field $A_{\mu}$ disappears. 
On the other hand, we have derived the generating functional in Eq. (\ref{23}) 
in a straightforward manner by taking the covariant gauges for 
$B_{\mu\nu}$ and $A_{\mu}@,@,$.  
By virtue of this method, origins of the two integrands in the 
exponent of Eq. (\ref{23}) have clearly been understood.

We now take  $J^{\mu\nu}$ to be the vorticity tensor current 
\cite{AE,Suga,ACPZ,KR}
\begin{eqnarray}
J^{\mu\nu}_{\varSigma}(x)\equiv 
2\pi Q \int_{\varSigma} d^{2}\xi @,@,@,@,@,
\frac{\partial X^{[@,@,\mu}(\xi)}{\partial \xi^{0}}
\frac{\partial X^{\nu@,@,]}(\xi)}{\partial \xi^{1}}@,@,@,@,@,@,
\delta^{(4)}@!@!@!@!@!\big(x-X(\xi)\big) \,,
\label{42}
\end{eqnarray}
%(42)
%
where $Q$ is a dimensionless coupling constant, 
$\xi=(@,@,\xi^{0}, \xi^{1})$ are the two-dimensional coordinates 
on the string world sheet, $\varSigma$, swept by the vortex string, 
and $X^{\mu}(\xi)$ denote the four-dimensional location of the
vortex string. 
When the vortex string has the end points, the tensor current 
$J^{\mu\nu}_{\varSigma}$ is no longer conserved, so that 
the non-zero vector current 
$j^{\mu}_{\partial\varSigma}(x)
\equiv\partial_{\nu} J^{\nu\mu}_{\varSigma}(x)
=-2\pi Q \int_{\partial\varSigma} dX_{\partial}^{\mu} @,@,@,
\delta^{(4)}(x-X_{\partial})$ occurs to satisfy Eqs. (\ref{4}) and (\ref{5}). 
Here $X^{\mu}_{\partial}$ stand for the locations of the end points. 
Substituting $j^{\mu}_{\partial\varSigma}$ into Eq. (\ref{23}) leads to 
the Yukawa potential, which is consistent with a result in Sec. 3.
On the one hand, substituting $J^{\mu\nu}_{\varSigma}$ into Eq. (\ref{23}) 
and carrying out the integration over $x^{\mu}$ lead to 
a non-local effective action written in terms of 
$\partial_{a} X^{\mu}(\xi)$ $(@,@,a=0,1 @,@,)$. 
After carrying out the Wick rotation, we can evaluate this effective action 
by the derivative expansion, obtaining the Nambu-Got\=o action as a leading 
term \cite{SY}. 
Furthermore, evaluating quantum fluctuation of $X^{\mu}(\xi)$ in the 
Nambu-Got\=o action, we find that the Nambu-Got\=o string 
yields the static linear potential at a long-distance scale \cite{Alv}. 
Therefore the composite of the Yukawa and the linear potentials 
is also derived from the generating functional in Eq. (\ref{23}) by 
choosing the vorticity tensor current $J^{\mu\nu}_{\varSigma}$ 
to be $J^{\mu\nu}$.

As we have seen in Sec. 2, the coupling term $mB_{\mu\nu}J^{\mu\nu}$ 
in Eq. (\ref{1}) leads to the non-local term 
$\dfrac{1}{4}m^{2}J_{\mu\nu}({\Box+m^{2}})^{-1}J^{\mu\nu}$ 
in the exponent of Eq. (\ref{23}).  
This term describes a dynamical vortex string, 
if $J^{\mu\nu}$ is taken as the vorticity tensor current in Eq. (\ref{42}). 
As has been mentioned above, the dynamical vortex string induces  
the linear potential at a long-distance scale.  
Thus we can say that the existence of 
a dynamical vortex string caused by the coupling term 
$mB_{\mu\nu}J^{\mu\nu}$ is essential for deriving the linear potential.

\begin{acknowledgements}
We are grateful to members of the Theoretical 
Physics Group at Nihon University for their encouragements. 
We would like to thank Dr. B. P. Mandal for a careful reading of 
the manuscript.
\end{acknowledgements}

\end{document}